\documentclass[11pt,twoside]{article}

\usepackage{asp2004}
\usepackage{psfig}
\usepackage{epsf}
\usepackage{lscape}
\usepackage{graphicx}
\begin{document}
\title{Magnetic Fields and UV-Line Variability in \boldmath $\beta$ Cephei}   
\author{R.S. Schnerr$^1$, H.F. Henrichs$^1$, S.P. Owocki$^2$, A. ud-Doula$^{2,3}$ and R.H.D. Townsend$^2$}   
\affil{
$^1$ Astron. Inst. ``Anton Pannekoek", Univ. of Amsterdam, Kruislaan 403, 1098 SJ Amsterdam, Netherlands\\
$^2$ Bartol Research Inst., Univ. of Delaware, Newark, DE 19716, USA\\
$^3$ Department of Physics, North Carolina State Univ., Raleigh, NC 27695-8202, USA 
}

\begin{abstract} 
We present results of numerical simulations of wind variability in the magnetic B1 IVe star $\beta$ Cephei. 
2D-MHD simulations are used to determine the structure of the wind. From these wind models we calculate line 
profiles for different aspect angles to simulate rotation. The results are compared with the observed UV wind 
line profiles.
\end{abstract}


\section{Introduction}
Winds of many early-type stars show cyclic, or even strictly periodic, variability on a rotational timescale. 
This behaviour was extensively studied in the UV with the IUE satellite. Both pulsations and magnetic fields 
have been suggested as the cause for this variability. It is clear that kG magnetic fields as observed in 
the Bp stars can explain the observed, strictly periodic, variability. For other early-type stars where no 
such fields have been detected and which can have much denser winds, the origin of the variability is still 
unknown. Recently, magnetic fields have been discovered in several early B type stars, with $\beta$ Cep 
among them. These stars have weaker fields \citep[a few hundred Gauss --][]{henrichs:2000a,coralie:2003a} 
and stronger winds than the magnetically strong Bp stars, which are mostly of a later spectral type.
Using numerical simulations, we investigate the effects of the magnetic field on the stellar wind of 
$\beta$ Cep.

\section{Method}
Our method consists of two steps. First we compute the wind geometry in the presence of a magnetic field, 
using a MHD code that starts out from the relaxed non-magnetic wind solution and a dipole field and is 
evolved in time until quasi-steady behaviour is observed. Second, from the end state of our simulations 
we calculate line profiles using a SEI (Sobolev with Exact Integration) method. Stellar rotation is 
simulated by adjusting the line-of-sight of the observer.

\begin{table}[t!bhp]
\begin{center}
\begin{minipage}[c]{8cm}
\caption{Stellar parameters of $\beta$ Cep}
\label{bcep_pars}
\end{minipage}\\
{\small
\begin{tabular}{ll|ll}
\tableline
spectral type   & B1IVe            &  polar field     & 360(30) G \\
mass            & 12 M$_{\odot}$   & rotation period & 12.00 days   \\
radius          & 6.5 R$_{\odot}$  & inclination     & ~60$\deg$--\,90$\deg$  \\
temperature     & $\sim$26000 K    & $\beta$         & ~90$\deg$  \\
\tableline
\end{tabular}
}
\end{center}
\end{table}

The {\bf 2D-MHD} code \citep[see][]{asif:2002} is based on ZEUS. It uses the standard MHD equations, the 
energy equation with a cooling term appropriate for stellar winds, and radial CAK-based radiative driving. 
Since $\beta$ Cep is a slow rotator, rotation can be ignored. For our models of $\beta$ Cep, we use basic 
stellar parameters as listed in Table~\ref{bcep_pars}. We treat mass loss rate, $\dot{M}$ as a parameter 
that ranges from $\sim$$10^{-8}$--$10^{-9}\, \mathrm{M_{\odot}\, yr^{-1}}$. The line profiles are calculated 
using the {\bf SEI} method. Opacities are determined using solar abundances and a (recently included) LTE ionisation balance. At low temperatures ($\sim \mathrm{T_{eff}}$) the CIV density 
is increased with a few percent due to the Auger effect. The highest relative CIV densities are found in 
shock heated sheets of $\sim$$10^5$ K. Our calculations do not account for the combined effect of the 
doublet members.

\begin{figure}[bt!]
\caption{Results of 2D-MHD simulations of the stellar wind in the presence of a 360 G dipole field and a 
mass-loss rate of the order of 10$^{-9}$ M$_{\odot}$ yr$^{-1}$. One can see the outflowing ``disk", in
 the magnetic equator, which has higher temperatures, higher densities and lower velocities than its 
surroundings.}
\label{snapshots}
\vspace{-0.4cm}
\begin{center}
\mbox{{\small $\log \rho$ (g)\hspace{0.2\textwidth} $\log$ T (K)}}\\
\end{center}
\vspace{-0.45cm}
\begin{minipage}{\textwidth}
\begin{center}
\includegraphics[width=0.32\textwidth]{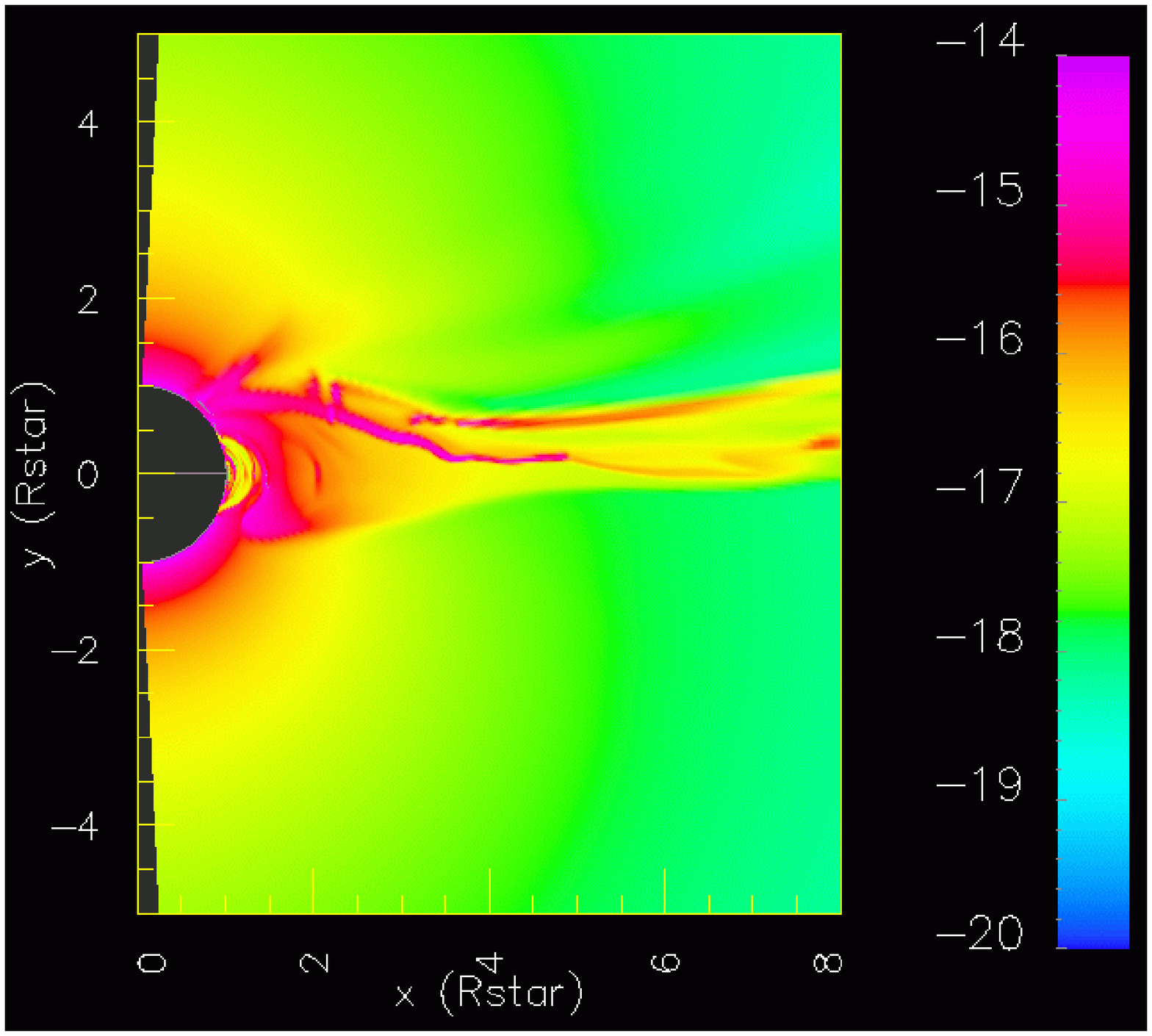}
\includegraphics[width=0.32\textwidth]{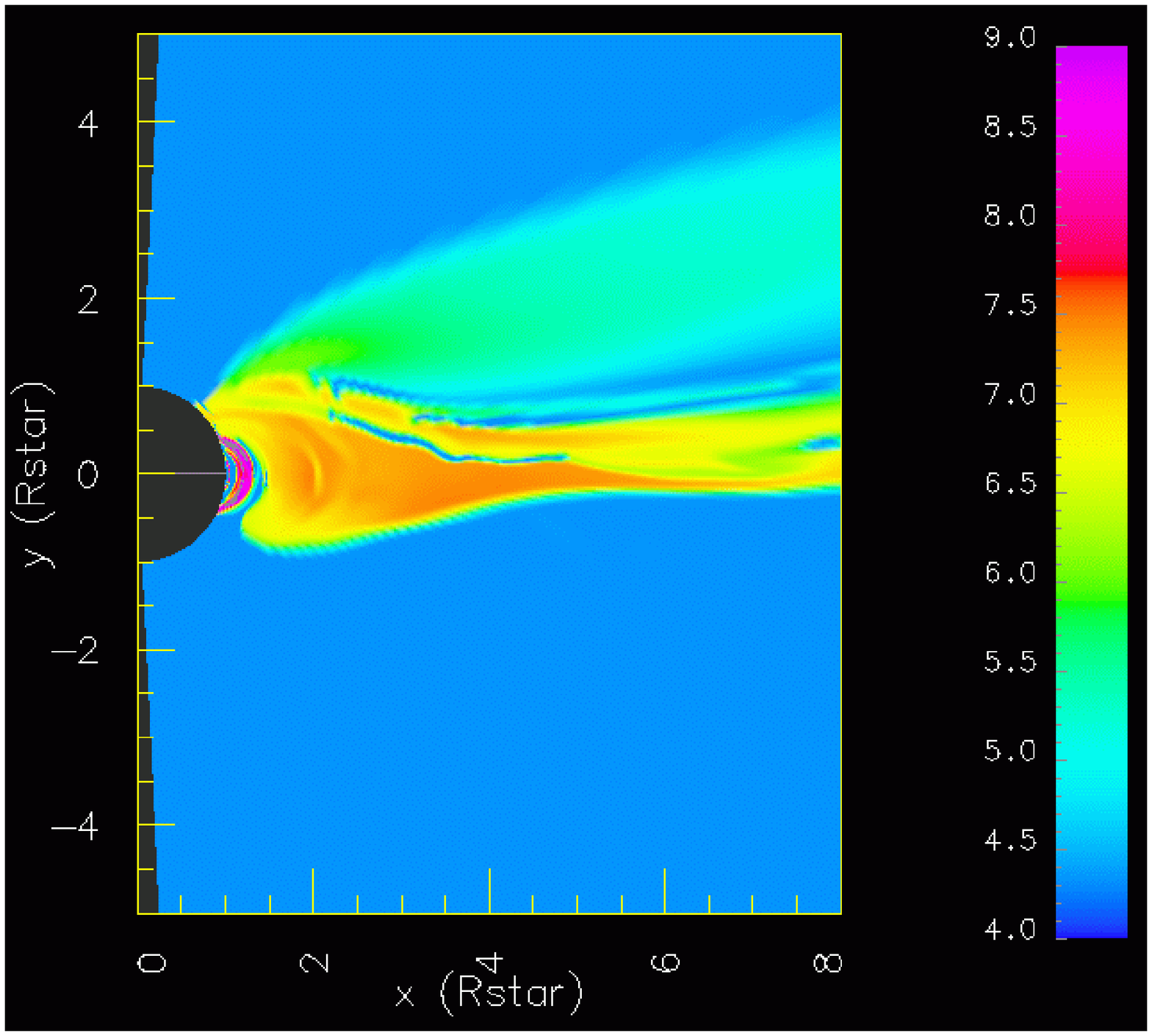}
\end{center}
\end{minipage}
\vspace{-0.4cm}
\begin{center}
\mbox{{\small $\mathrm{v_{rad}\ (km\,s^{-1})}$ \hspace{0.2\textwidth} $\mathrm{v_{\theta}\ (km\,s^{-1})}$}}\\
\end{center}
\vspace{-0.35cm}
\begin{minipage}{\textwidth}
\begin{center}
\includegraphics[width=0.32\textwidth]{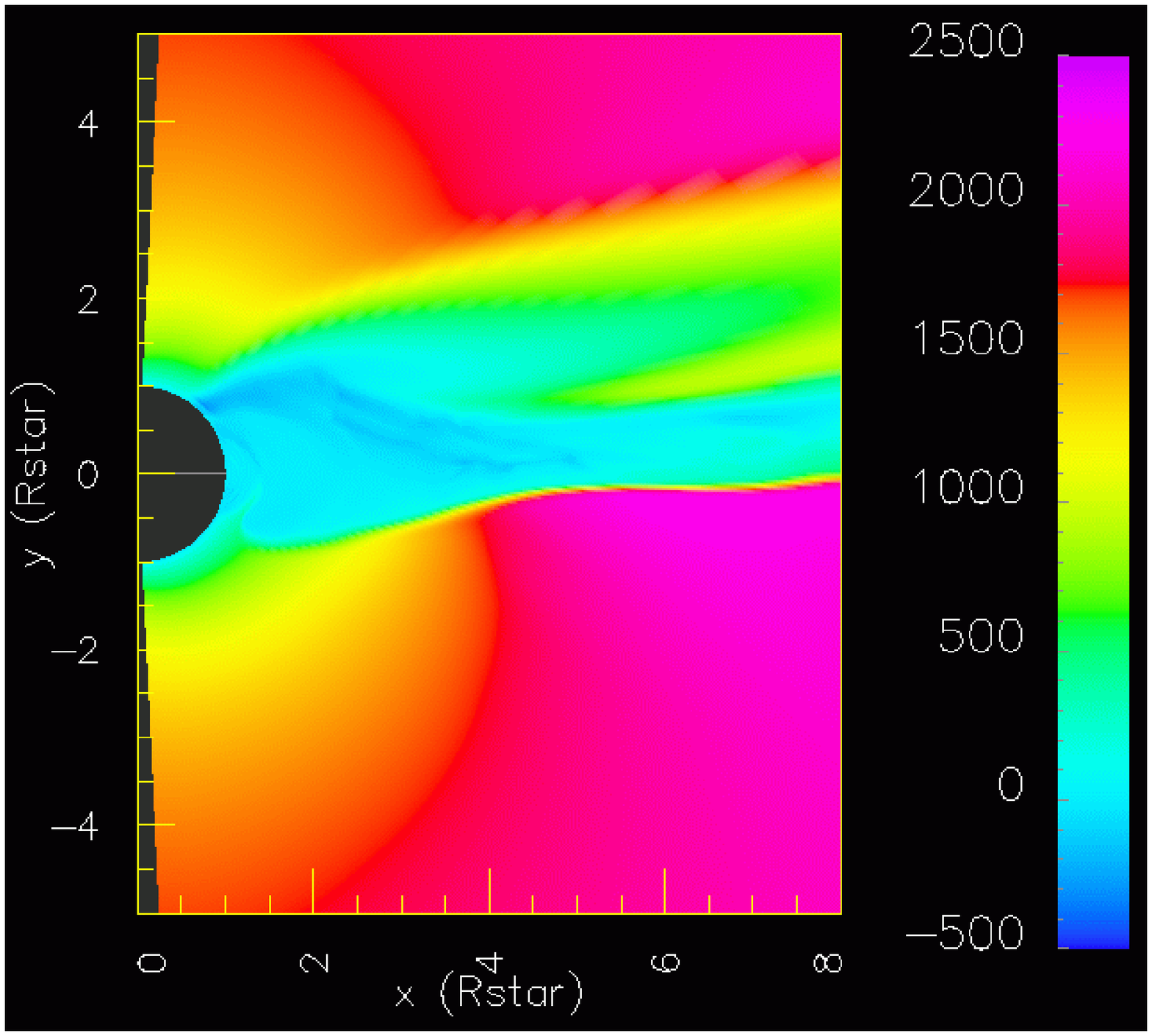}
\includegraphics[width=0.32\textwidth]{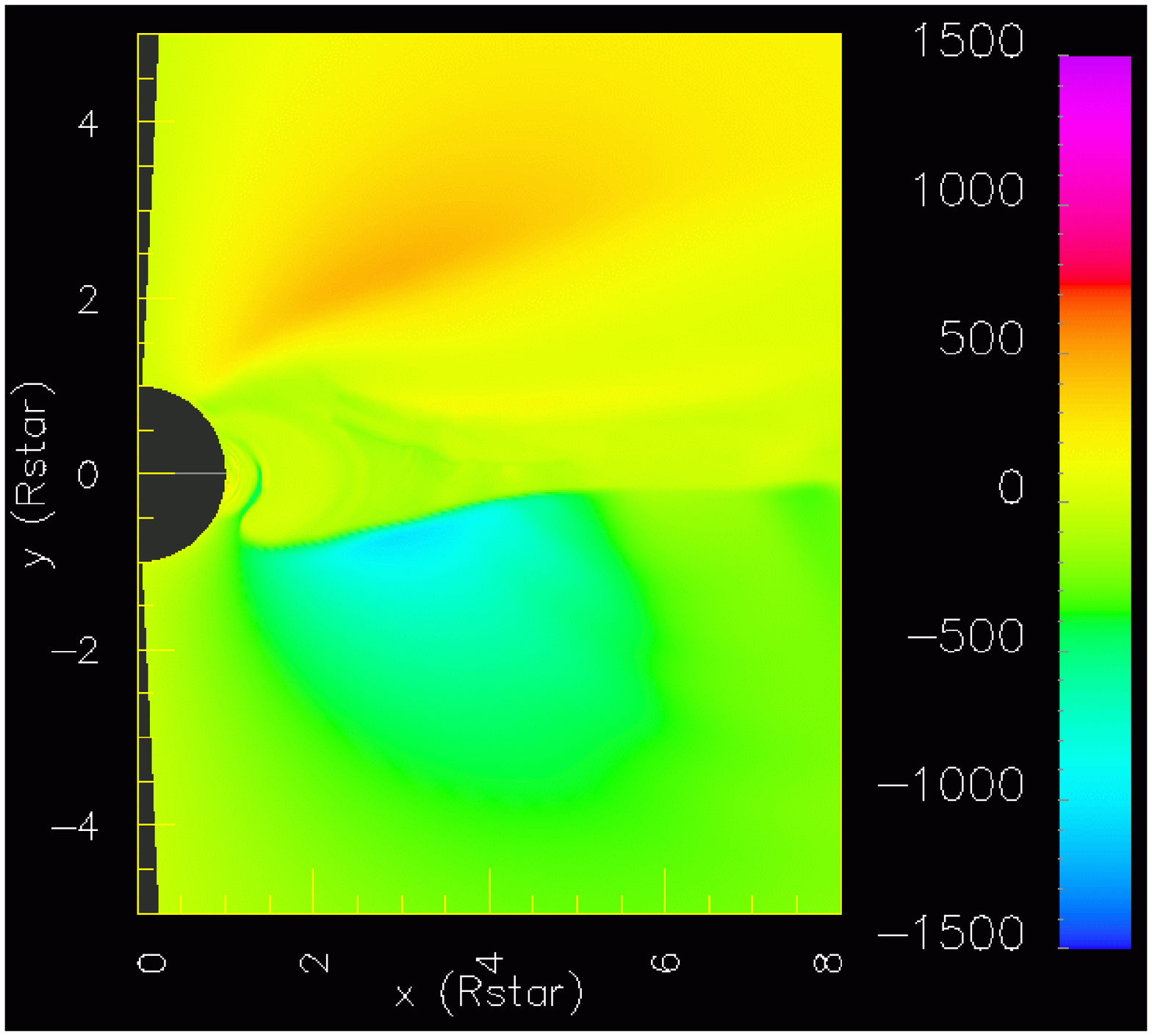}
\end{center}
\end{minipage}
\end{figure}

\begin{figure}[p!htb]
\caption{{\bf (Left)} Strictly periodic UV wind-line variability as observed in the CIV doublet of $\beta$~Cep. 
The top panel shows the significance of the variability. {\bf (Middle)} Results for a model with 
B$_\mathrm{p}$=190 G and a fixed CIV abundance. The basic features are reproduced, but fine-tuning of the 
parameters is needed. {\bf (Right)} Results for B$_\mathrm{p}$=360 G, including a temperature dependent CIV 
abundance. Note the 0.25 offset in phase relative to the observations.}
\label{lineprofiles}
\includegraphics[width=0.325\textwidth, trim=0 0 5 10, clip]{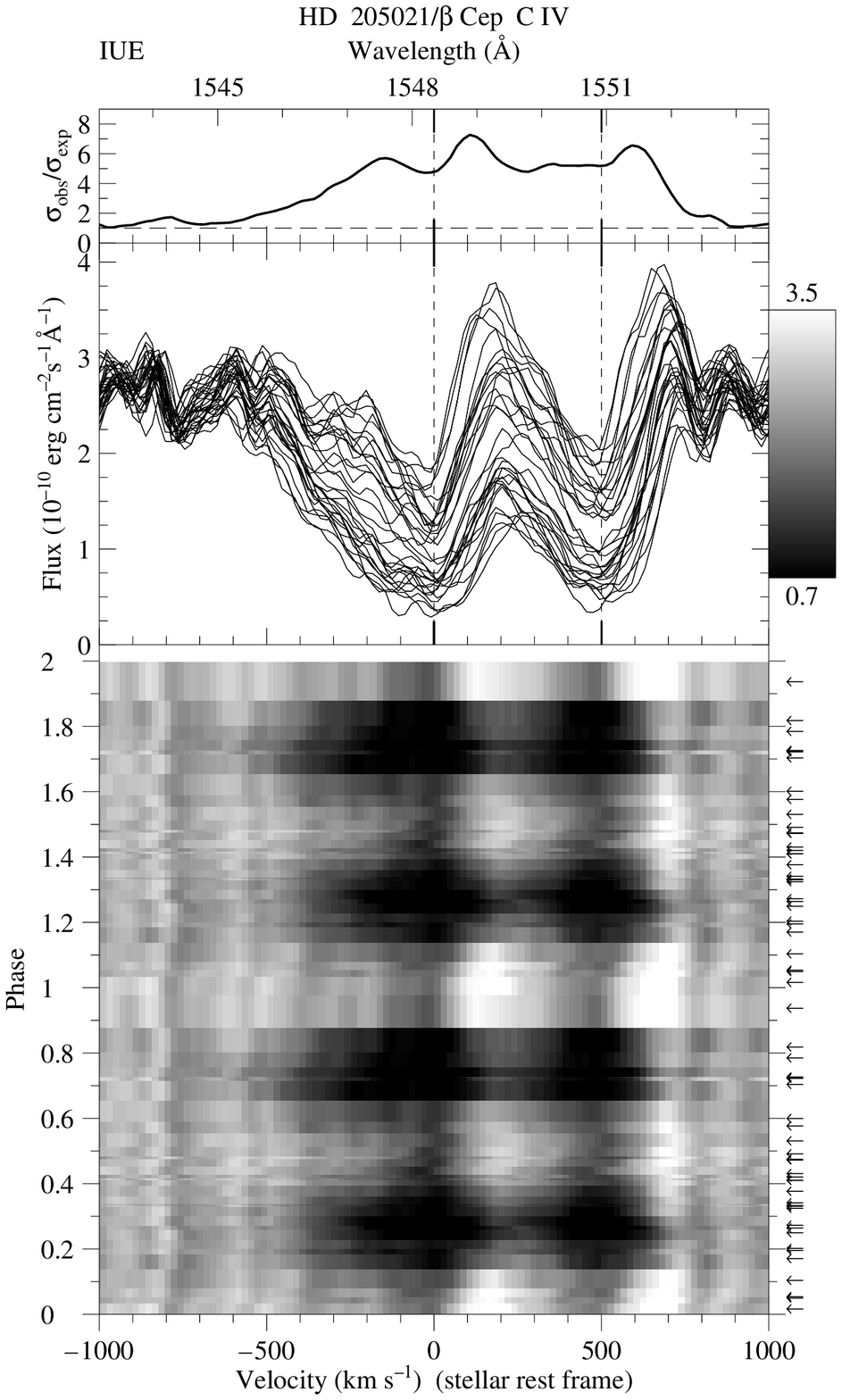}
\includegraphics[width=0.325\textwidth, trim=0 0 5 10, clip]{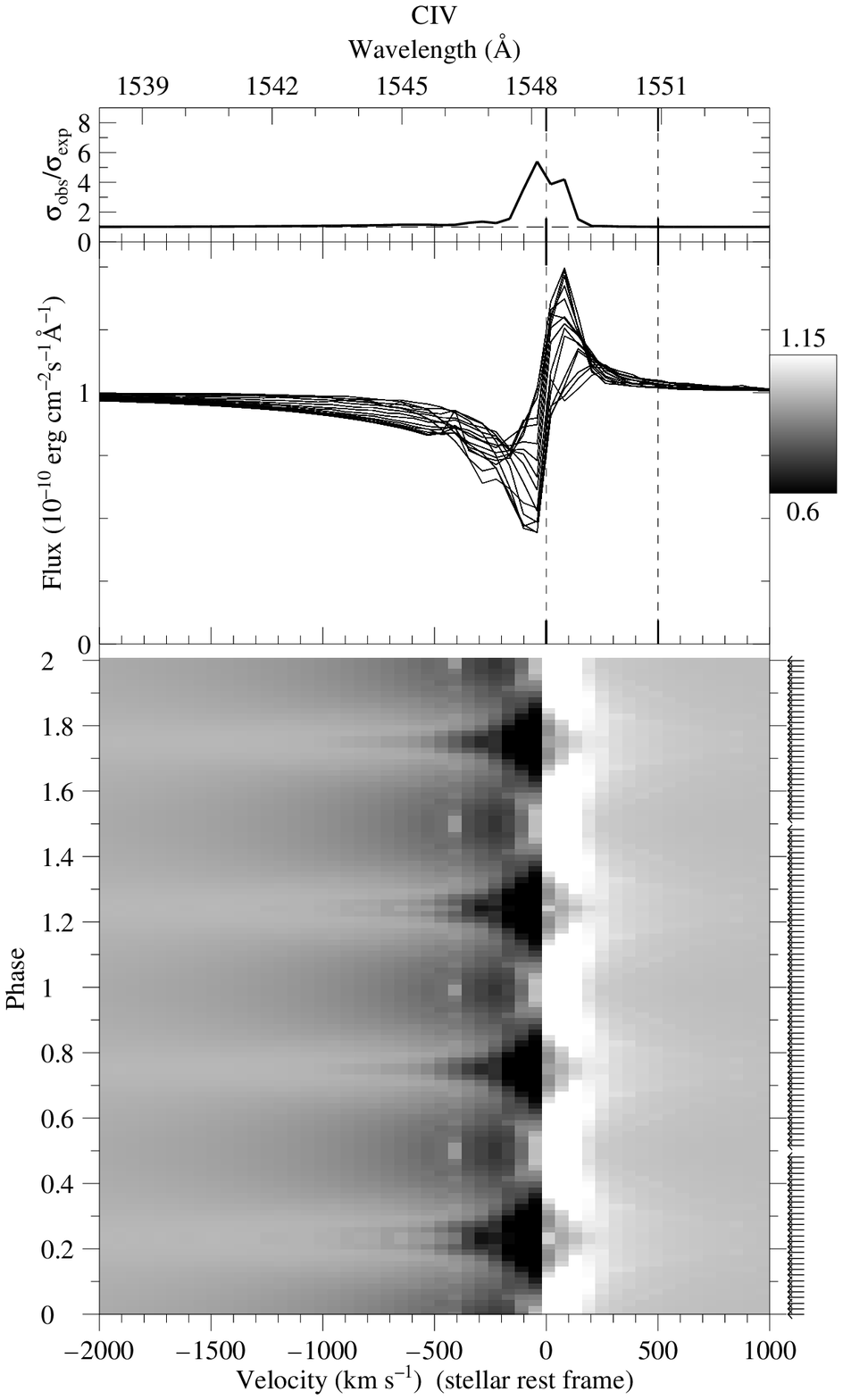}
\includegraphics[width=0.325\textwidth, trim=0 0 5 10, clip]{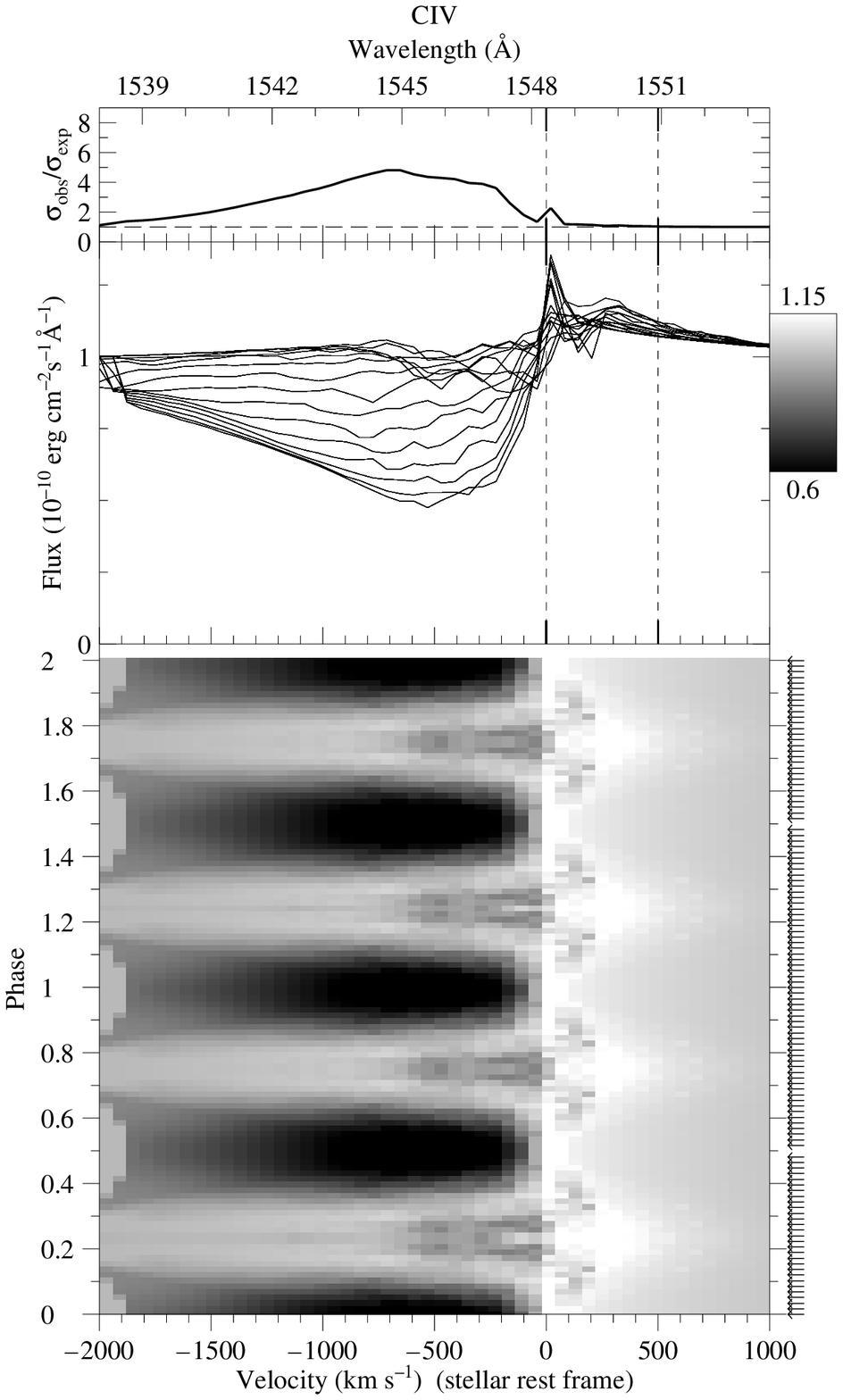}
\end{figure}

\section{Results and Conclusions}
For a model with a polar field ($\mathrm{B_p}$) of 360 G and mass loss rate of 
$\sim$10$^{-9}$ M$_{\odot}$ yr$^{-1}$, Fig.\,\ref{snapshots} shows the density, temperature, and velocity at 
a typical time snapshot 
long after any influence of the initial state has faded. The material is guided by the magnetic field towards 
the magnetic equator, where it is heated by shocks. Due to the higher densities in this ``disk'', the radiation force is reduced and radial velocities are relatively small.
In Fig.\,\ref{lineprofiles} we show the observed CIV variability as a function of the rotational phase and calculations
for the blue line of the doublet. Both in the observations and the 
simulations with a fixed abundance, most absorption is seen along the magnetic equatorial plane. For the 
simulations including the ionisation balance, the pattern appears to be offset by 0.25 in phase. This is 
caused by the high temperature in the disk which reduces the CIV abundance. A more detailed treatment of the 
Auger effect is likely required to account for this offset, since most X-rays will come from the shock heated 
disk. 
Recent results \citep{leone:2005} show that magnetic fields measured with the LSD method could
be overestimated. If $\beta$ Cep would indeed have a weaker field, this could help to better explain the observed wind terminal speed.
Although more work needs to be done, in principle the effect of the magnetic field on the wind of $\beta$~Cep 
can explain the observed UV line-profile variability. Further modelling of other magnetic early B stars and 
wind lines like NV and SiIV, will allow a detailed view on how stellar winds are affected by magnetic fields, 
and the importance of the Auger effect.



\begin{thebibliography}{}
\bibitem[Henrichs et al.(2000)]{henrichs:2000a}
Henrichs, H.~F., de~Jong, J.~A., et al. 2000, in ASPCS Vol\,214, The Be phenomenon in 
  early-type stars, ed.\ M.A. Smith, H.F. Henrichs and J. Fabregat, p.\,324
\bibitem[Leone et al.(2005)]{leone:2005}
    Leone, F., Schnerr, R.~S., Stift, M.~J., \& Henrichs, H.~F. 2005, \aap, in prep.
\bibitem[Neiner et al.(2003abc)]{coralie:2003a}
    Neiner, C., Geers, V.~C., Henrichs, H.~F., et~al. 2003a, \aap, 406, 1019
\bibitem[Neiner et al.(2003b)]{coralie:2003b}
    Neiner, C., Hubert, A.-M., Fr{\' e}mat, Y., et~al. 2003b, \aap, 409, 275
\bibitem[Neiner et al.(2003c)]{coralie:2003c}
    Neiner, C., Henrichs, H.~F., Floquet, M., et~al. 2003c, \aap, 411, 565
\bibitem[ud-Doula \& Owocki(2002)]{asif:2002}
    ud-Doula, A.,  \& Owocki, S.~P. 2002, \apj, 576, 413
\end{thebibliography}
\end{document}